\documentstyle[12pt,epsfig]{article}
\input epsf

\large
\newcommand{\be}{\begin{eqnarray}}
\newcommand{\ee}{\end{eqnarray}}

\newcommand{\mat} {\left ( \begin{array}{cc}}
\newcommand{\emat} { \end{array}\right )}

\begin{document}
\setlength{\baselineskip}{17pt}
\pagestyle{empty}
\vfill
\eject
\begin{flushright}
SUNY-NTG-00/12
\end{flushright}

\vskip 2.0cm
\centerline{\Large \bf  Spectral Universality of Real Chiral}
\vskip 1cm
\centerline{\Large \bf Random Matrix
Ensembles}
\vskip 1.2cm
\centerline{ B. Klein and J.J.M. Verbaarschot}
\vskip 0.2cm
\centerline{\it Department of Physics and Astronomy, SUNY, 
Stony Brook, New York 11794}
\vskip 0.2cm
\centerline{\tt klein@grad.physics.sunysb.edu,\,\,\, 
verbaarschot@nuclear.physics.sunysb.edu}
\vskip 1.5cm

\centerline{\bf Abstract}
We investigate the universality of microscopic eigenvalue correlations
for Random Matrix Theories with the global 
symmetries of the QCD partition function. In this article we analyze the
case of real valued chiral Random Matrix Theories 
($\beta =1$) by relating
the kernel of the correlations functions for $\beta =1$
to  the kernel
of chiral Random Matrix Theories with complex matrix elements ($\beta = 2$),
which is already known to be universal. Our proof is based on 
a novel asymptotic property of the skew-orthogonal polynomials: 
an integral over the corresponding wavefunctions 
oscillates about  half its 
asymptotic value in the region of the bulk of the zeros. 
This results solves the puzzle that microscopic universality persists 
in spite of contributions to the microscopic correlators
from the region near the largest zero of the skew-orthogonal polynomials. 
Our analytical results are illustrated by the numerical construction of
the skew-orthogonal polynomials for an $x^4$ probability potential.
 
\vfill
\noindent

\eject
\pagestyle{plain}

\section{ Introduction}

Since its first application to the spacing of nuclear 
resonances \cite{Wign51}, 
Random Matrix Theories (RMT) have been very successful in explaining the
statistical properties of spectra. Originally, the so called
Wigner-Dyson ensembles of matrices with independently distributed
Gaussian matrix elements
were introduced to replace the  unknown nuclear Hamiltonian.
 More recently, the applicability of RMT has been related to  the
chaotic dynamics of the corresponding classical system and spectra
of many chaotic systems with only a few degrees of freedom have been
successfully described by RMT \cite{bohigas,Seli84,HDgang}. 
The successes of RMT has raised the question whether
the statistical properties of spectra are universal. This question
has been investigated in great detail within the context of RMT. The 
idea is to show that large deformations of the probability distribution
of the random matrix elements leave the properly
rescaled spectral correlations unaffected, whereas the average spectral
density changes on a macroscopic scale.
This program has been carried out most completely for the Hermitian 
random matrix ensembles 
(denoted by the Dyson index $\beta =2$; for recent reviews
 see \cite{HDgang,Kanz97})
which are mathematically much simpler than
real or quaternion-real random matrix ensembles (with Dyson index
$\beta = 1 $ and $\beta = 4$, respectively). 
Nevertheless,
several universality proofs are available for 
these ensembles as well 
\cite{mm,BN,Been93,Hack95,Senerprl,Widom,akerel,Toub99,akekan}.

In addition to the  Wigner-Dyson ensembles there are seven other 
classes of Random Matrix Theories. They can be classified according
to the Cartan classification of symmetric spaces \cite{class}.
In this article we are interested in chiral Random Matrix Theories (chRMT).
These are ensembles of random matrices with the chiral
symmetry of the QCD Dirac operator \cite{SVR,V}. 
The nonzero eigenvalues of these  ensembles  occur
in pairs $\pm \lambda$. Therefore, $\lambda =0$ is a special 
point, and  the average spectral density on the scale of the average
level spacing  shows universal properties. 
With the average spacing of the eigenvalues given by $\pi/\Sigma N$
(with $N$ the total number of eigenvalues and $\Sigma$ a parameter
known as the chiral condensate),
the microscopic spectral density is defined as \cite{SVR}
\be
\rho_s(u) = \lim_{N\rightarrow \infty} \frac 1{\Sigma N} \langle
\rho(\frac u{ \Sigma N})\rangle.
\label{rhosu}
\ee  
Both $\rho_s(u)$ and the microscopic $k$-point correlation functions 
are universal. This has
been shown in great detail 
for the chiral Unitary 
Ensemble (chUE), which is the ensemble of Hermitian chiral random 
matrices with no anti-unitary symmetries
\cite{AJ,ADMN,brezin,kanz,Kanz97,Sener1,GWu,Seneru,splan,spli,Damg97,Tilodam}
\cite{massuni,akerel,Damg98a,OTV,DOTV,Seif}.
The chiral ensembles with real
or quaternion real matrix elements 
(known as the chiral Orthogonal Ensemble (chOE) and the chiral 
Symplectic Ensemble (chSE), respectively)
are mathematically much more complicated. The general result for
the microscopic spectral  
density of the chiral Gaussian Orthogonal Ensemble (chGOE) \cite{V2} and the
of  the chiral Gaussian Symplectic Ensemble (chGSE) 
\cite{Nagao-Forrester} was first obtained by an explicit
construction of the corresponding 
skew-orthogonal polynomials \cite{Dyson-skew,mm,nagao}
(several special cases were analyzed in \cite{nagaochi,brezin,Altland}).
This method does not seem to be easy generalizable to non-Gaussian 
probability potentials.

Recent progress was made by relating the kernel for the correlation
functions of the chOE to the universal kernel 
of the chUE \cite{Senerprl}. 
This method was based on 
a generalization of an operator
construction of  skew-orthogonal polynomials for the
Wigner-Dyson ensembles \cite{BN} to the chiral ensembles.
Remarkably, the skew-orthogonal 
polynomials do not enter in the relation between the kernels.
Indeed, an elegant construction of the correlation functions relying only
on operator relations was recently given in \cite{Senerprl,Widom}. 
The operator method was successfully applied to  Gaussian Orthogonal
and Gaussian Symplectic Ensembles with an additional fermion determinant
\cite{hilmoine}.  
Universality for the so called massive chiral ensembles for $\beta = 1$ and
$\beta = 4$ was shown by relating them to the corresponding massless
ensembles \cite{akekan}. In this way analytical results 
 for the massive spectral correlators could be obtained
\cite{akekan,Nishim}.

 The universality proof based on the relations between kernels 
given in \cite{Senerprl} is incomplete for $\beta =1$. The reason is
that the correlation functions for $\beta =1$ 
depend on a integral over the positive real axis of a derivative of the kernel 
for $\beta = 2$. Naively, one would
expect that the presence of such nonlocal contributions
 would lead to nonuniversal
behavior. The resolution of this puzzle, 
and thus the completion of the universality proof for $\beta =1$, 
is the primary objective of this paper.
As was already discussed in \cite{Senerprl}, for $\beta =4$ there is no
such problem. As secondary objective, we illustrate
some of the arguments given in \cite{Senerprl} 
by the analysis of the asymptotic behavior of the 
skew-orthogonal polynomials for a quartic probability potential. 

The results of this article are relevant for Dirac spectra 
of QCD with two colors in the fundamental representations
(or for Dirac spectra of 
QCD with staggered fermions in the adjoint representation). 
Indeed, the microscopic spectral density of the chGOE has been
observed 
in lattice QCD \cite{heller,Edwa99b}
and in instanton liquid simulations \cite{Vinst}. Many more
results justifying the chiral Random Matrix description of the
 microscopic spectral density in lattice QCD have been
obtained for $\beta =2$ and $\beta =4$ \cite{Vplb,Tilo,Damg99a,Gock98,Farc99a}
(for a recent review and a complete list of references see \cite{TVrev}).
  Our results also apply  to 
the superconducting ensembles with Dyson index $\beta = 1$
(with joint eigenvalue density given
by a special case of the chiral ensembles) as well as to two-sublattice
models \cite{Gade,brezin}.

This paper is organized as follows. Chiral Random Matrix Theory is
introduced in section 2. In section 3 we discuss the relation
between the kernels for $\beta = 1$ and $\beta=2$. Most of this section
already appeared in \cite{Senerprl}. Two examples,
the Gaussian case and the quartic probability potential are
worked out in detail in section 4. 
In section 5 we derive a novel asymptotic property of the
skew-orthogonal polynomials, allowing us to complete the proof of 
\cite{Senerprl}. This is the most important result of this paper.
Universality of the microscopic correlations is
shown in section 6. In this section we also give the explicit 
universal expressions for
the microscopic spectral density and the microscopic kernel which
allows to  calculate all correlation functions. In section 7 
the skew-orthogonal 
polynomials for a quartic probability potential are studied numerically  
and  concluding remarks are made in section 8. 

\section{Chiral Random Matrix Theory}

In this section we introduce a Random Matrix Theory with the global
symmetries of the QCD partition function. We first define the 
Dyson index $\beta$ of a Dirac operator. Its value is equal to
the number independent degrees of freedom per matrix element and
is determined by the anti-unitary symmetries of the Dirac operator.
Any anti-unitary symmetry operator can be written as
$U = AK$ with $A$ unitary and $K$ the complex conjugation operator. 
The operator $U^2$ is unitary and is thus proportional to the identity
in an irreducible subspace of the unitary symmetries of the Dirac operator.
One can easily convince one-self that the only two possibilities 
are $A^2 = 1$ and $A^2 = -1$ in this subspace. These two cases are denoted
by the Dyson index $\beta = 1$ and $\beta=4$, respectively. If there
are no anti-unitary symmetries, the Dyson index of the Dirac operator
is $\beta = 2$. For $\beta = 1$ it is possible to find a basis for which
the Dirac matrix is real for {\it all} gauge field configurations. 
For $\beta =4$
it is possible to construct a basis for which the Dirac matrix
is quaternion real for {\it all} gauge field configurations. For $\beta =2$
the matrix elements of the Dirac operator do not have any reality properties
and are given by complex numbers.
The Dyson index of the Dirac operator depends on the representation of
the gauge fields and  may be different for the discretized and continuum 
versions of
the Dirac operator \cite{V,Halasz}.  For example, the continuum
Dirac operator for $N_c = 2$ and  fermions in the fundamental representation
is in the class $\beta =1$, whereas 
the corresponding staggered Dirac operator is in the class $\beta =4$.
For gauge fields in the adjoint representation the Dyson index of the
continuum Dirac operator is $\beta =4$, whereas the Dyson index of
the staggered lattice discretization is $\beta =1$. 

In a chiral basis in the sector of topological charge $\nu$ the
Dirac matrix has the block structure
\be
D = \mat 0 & C \\ C^\dagger & 0 \emat,
\ee
where $C$ is an $n \times (n+\nu) $ matrix with reality properties given
by the Dyson index. For generic values of its matrix elements 
D has exactly $\nu$ zero eigenvalues. 
In QCD, the matrix elements of $D$ depend in a complicated way on the gauge
fields, which are distributed according to the QCD partition function. 
In chiral Random Matrix Theory (chRMT) we replace the matrix elements of the
Dirac operator by space-time independent random numbers with 
probability distribution given  by the partition function
\be
Z(m) = m^\nu \int d C {\det}^{N_f}( D + m) e^{-\frac { n\beta}2 
{\rm Tr} V(C^\dagger C)}.
\label{part}
\ee 
Here, $N_f$ is the number of quark flavors with mass $m$ (below
we only consider the massless case $m=0$). 
The integral is over the independent degrees of freedom of the matrix
elements of $C$. The probability potential $V(x)$ is in general an arbitrary
polynomial $V(x)= \sum_{k=1}^p a_k x^k$ (with $a_p > 0$). 
For the Gaussian chiral ensembles,
which are mathematically much simpler, 
the potential is $V(x) = \Sigma^2 x$. In that case  
the parameter  $\Sigma$ is related to the average spectral density
\be
\rho(\lambda) = \langle \sum_{k=1}^N \delta(\lambda -\lambda_k)\rangle 
\ee
 by  the Banks-Casher formula \cite{BC}
\be
\Sigma = \lim_{N\rightarrow \infty}\frac{\pi \rho(0')}N.
\ee
The prime indicates that the argument of
$\rho(\lambda)$ should be near zero but much larger than the smallest
eigenvalue. For this reason $\Sigma$ is interpreted as the chiral
condensate, the order parameter of the chiral phase transition.
The integrals in (\ref{part}) can be easily evaluated in the thermodynamic
limit. The result coincides with  so called 
finite volume partition functions which were first derived on the basis
of chiral symmetry \cite{LS,SV}.

In this paper we will study the chiral Orthogonal Ensemble (chOE) and
show that the microscopic spectral density does not depend
on the coefficients of the probability potential. Our starting
point is the joint probability distribution of the eigenvalues
of the Dirac operator. It can be obtained form the partition function
by making a polar decomposition of $C$
\be
C = U \Lambda V^{-1},
\ee
with $U$ and $V$ orthogonal, unitary or symplectic matrices for
$\beta =1$, $\beta =2 $ and $\beta =4$, respectively, and $\Lambda$ a
semi-positive definite diagonal matrix.
For the massless case the joint probability distribution in
terms of the squares of the eigenvalues, $x_k= \Lambda_k^2$, is given by
\be
\rho(x_1, \cdots, x_n) = 
  |\Delta(\{ x_i \})|^\beta
  \prod_k x_k^{N_f-1 +\beta|\nu|/2+\beta/2} 
  e^{- \frac{n\beta}2 \sum_k V(x_k)}
  \label{zeig1}
\ee
where the Vandermonde determinant is defined by
\be
  \Delta(\{x_i \}) = \prod_{k<l} (x_k-x_l)\:.
  \label{vandermonde}
\ee   
The exponent of the $x_k$ for $\beta =1$ will be denoted by
\be
a=N_f-1 +|\nu|/2+1/2 ,
\ee 
and below we consider the joint probability density
\be
\rho(x_1, \cdots, x_n) = 
  |\Delta(\{ x_i \})|^\beta
  \prod_k e^{-\beta \phi_a(x_k)}
  \label{zeig}.
\ee
with probability potential given by 
\be
\phi_a = \frac n2 V(x) -a\log x.
\ee
For technical reason we will restrict ourselves to even $n$ and
use the notation $n= 2\bar n$.

The spectral density, and in general the $k$-point correlation
functions,  can be
obtained from the joint eigenvalue density by integrating over
all but $k$ eigenvalues. 
For $\beta = 2$ this can be simply done by exploiting the 
orthogonality of the polynomials \cite{Mehta}
\be
\int_0^\infty dx e^{-2\phi_a(x)} P_k^{2a}(x) P_l^{2a}(x) =\delta_{kl}.
\ee
The resulting spectral correlation functions can be expressed in terms of
the kernel\footnote{For later convenience we do not follow the usual
convention to include the weight functions in the kernel.}
\be
K^{2a}_{n}(x,y) = 
\sum_{k=0}^{n-1} P_k^{2a}(x) P_k^{2a}(y). 
\ee
Universality can be established by showing that 
in the microscopic limit the so called wave functions 
$P_k^{2a}(x)\exp(-\phi_a(x))$ depend only
on the potential through a scale factor determined by the average
spectral density \cite{ADMN}.
With $\Sigma = \pi \rho(0)/2n$ 
the microscopic limit of the kernel for the chUE is given by
(a factor $2 \sqrt {uv}$ from the integration measure has been included)
\be
\lim_{N=2n\to \infty} \frac 2{\Sigma N} 
 \left ( \frac{ {uv}}{\Sigma^2 N^2}\right )^{2a+1/2}
 K_n^{2a}(
\frac{u^2}{\Sigma^2 N^2},\frac{v^2}{\Sigma^2 N^2})
&=&\sqrt{uv} \:\frac{uJ_{2a+1}(u)J_{2a}(v)
    -vJ_{2a}(u)J_{2a+1}(v)}{u^2-v^2}\nonumber \\
&\equiv& B^{2a}(u,v).
\nonumber \\
\label{besselkern}
\ee
This kernel is known as the Bessel kernel \cite{Tracy}. Sometimes it is
simpler to use an integral representation of the Bessel kernel given by
\be
B^{2a}(u,v)= \sqrt{uv} \int_0^1 t dt J_{2a}(ut) J_{2a}(vt).
\label{besselkernint}
\ee

\section{Relation between the kernel for $\beta=1$ and $\beta=2$}
\label{relation}

For the orthogonal ensembles the integrations over the eigenvalues
can be performed
by means of orthogonality relations for the skew-orthogonal polynomials
of the second kind \cite{Dyson-skew,mm}. 
These polynomials  are defined by the scalar products
\be
\langle R_k, R_l \rangle_R = J_{kl},
\label{orthoskew}
\ee
with the nonzero matrix elements of $J_{kl}$ given by $J_{2k,2k+1} =
-J_{2k+1,2k} = -1$. The skew-scalar product is defined by
\be
\langle f, g \rangle_R = \int_0^\infty dx e^{-2\phi_a(x)}
f(x) \hat Z g(x),
\ee
where we have introduced the operator $\hat Z$ by \cite{BN,Senerprl},
\be
\hat Z g(x) = \int_0^\infty dy e^{\phi_a(x)} 
\epsilon(x-y) e^{-\phi_a(y)} g(y).
\ee
 As usual, $\epsilon(x) = x/2|x|$.
All correlation functions can be expressed in terms of the kernel \cite{mm}
\be
K_R^a(x,y) = \int_0^x dze^{-\phi_a(z)} k_R^a(y,z) e^{-\phi_a(y)},
\label{KR}
\ee
with the pre-kernel defined by
\be
k_R^a(y,z) = \sum_{i,j=0}^{2\bar n-1} R_i^a(y) J_{ij} R_j^a(z).
\ee
For example, the spectral density is given by
\be
\rho(x) = K_R^a(x,x) - \frac 12 K_R^a(\infty,x).
\label{spect}
\ee
 
We construct the skew-orthogonal polynomials for the chUE by a generalization
of an operator method \cite{Senerprl}
introduced by Br\'ezin and Neuberger \cite{BN}.
The skew-orthogonal polynomials are expressed in  terms of orthogonal
polynomials $P_k^{2a+b}$ with weight function $x^{2a+b}\exp(-nV(x)/2)$,
\be
R_i^a(x) = \sum_{j=0}^i T_{ij} P_j^{2a+b}(x).
\ee
We will derive recursion relations for the expansion coefficients $T_{ij}$.
It is useful to  introduce
the operators
\be
\hat X, \quad \hat X^b \hat \partial, \quad \hat X^{-b} \hat Z,
\label{op3}
\ee
and the operator
\be
\hat L = \hat X^b \hat \partial - \hat X^b \phi'(\hat X) + b \hat X^{b-1}.
\label{opl}
\ee 
The coordinate operator and the derivative operator are defined by
$\hat X f(x) = x f(x)$ and $\hat \partial f(x) = f'(x)$, respectively,
and the operator $\hat L$ is the inverse of $\hat X^{-b} \hat Z$, i.e.,
\be
\hat L \hat X^{-b} \hat Z g(x) = g(x).
\ee
The matrix elements of the operators in (\ref{op3}) and (\ref{opl})
with respect to the basis  $P_k^{2a+b}(x)$ will be denoted by
$X_{kl}$, $D_{kl}$, $Y_{kl}$ and $L_{kl}$, in this order.
For an operator $\hat A$ the matrix elements are defined by
\be
A_{kl} = \int_0^\infty dx e^{-2\phi_a(x) + b\log x}  P_l^{2a+b}(x) 
\hat A  P_k^{2a+b}(x).
\ee
By partial integration it can be shown that  $L$ and
$D$ are related by
\be
L_{kl} = \frac 12(D_{kl} -D_{lk}).
\ee
For integer values of{\,}\footnote{For $2a = 0$ the matrix
$L$ remains a band matrix  even for $b=0$.} $b \ge 1$ 
the matrix elements of $L$ 
thus vanish for  $|k-l| > {\rm max} (b-1,b+ p-1)$
(where $p$ is the order of the polynomial probability potential). 
The optimum value of $b$ is  thus $b=1$ which will be our choice in
the remainder of this article.

The orthogonality relation (\ref{orthoskew}) can be written as
\be
TY T^T = -J.
\ee
By acting with $L$ and $T^T JT $ on $YT^T$ it follows that
\be
L = T^T J T.
\label{tjt}
\ee
In matrix form this equation can be rewritten as
\be
L_{kl} = \sum_{p=0}^\infty [T_{2p+1,k} T_{2p,l} - T_{2p,k}T_{2p+1,l}].
\label{lexplit}
\ee
For known $L_{kl}$  the coefficients $T_{ik}$ can be determined
recursively from this relation. If the $L_{kl}$ vanish outside a band
$|k-l| > p$ (as is the case for $V(x)$ given by a polynomial
of order $p$), the coefficients $T_{ik}$ are nonzero
only inside a  band $i-k< M <2p$. 

The pre-kernel can be written as
\be
k_R^a(x, y) &=& \sum_{i,j=0}^{2\bar n-1} \sum_{k \leq i} \sum_{l\leq j}    
P_k^{2a+1}(x)T^T_{ki}J_{ij}T_{jl}P^{2a+1}_l(y)\nonumber \\
&=&\sum_{k,l=0}^{2\bar n-1} \sum_{i,j}     
P_k^{2a+1}(x)T^T_{ki}J_{ij}T_{jl}P^{2a+1}_l(y)
- R(x,y),
\label{pret}
\ee
where remainder term is given by
\be
R(x,y) =  \sum_{i,j=2\bar n}^{2\bar n+M-2} \sum_{k \leq i} \sum_{l\leq j}    
P_k^{2a+1}(x)T^T_{ki}J_{ij}T_{jl}P^{2a+1}_l(y).
\ee
There are now no restrictions on the  summation  over $i$ and $j$ in 
the last line of (\ref{pret}) and the relation (\ref{tjt}) can be used 
to simplify the expressions. For finite $M$ and a smooth dependence
of the coefficients $T_{ij}$ on the order of the skew-orthogonal polynomials
the number of terms in $R(x,y)$, and thus the contribution of
$R(x,y)$ to the pre-kernel, is subleading in $1/n$. In the next section, 
this will be shown explicitly
both 
for a Gaussian and a quartic probability potential.
To leading order in $1/n$ we thus find
\be
k_R^a(x, y) &=& \sum_{k,l=0}^{2\bar n-1} 
P_k^{2a+1}(x)L_{kl}P^{2a+1}_l(y)\nonumber \\
&=&\frac 12 \sum_{k,l=0}^{2\bar n-1} 
P_k^{2a+1}(x)[D_{kl} - D_{lk}]P^{2a+1}_l(y).
\label{pred}
\ee
By re-expressing the matrix elements of $D$
in terms of the operators $x\partial_x$ and $y\partial_y$ we
find the following relation between the kernel for the chiral Orthogonal
Ensemble and the kernel for the chiral Unitary Ensemble \cite{Senerprl}
\be
k_R(x, y) = \frac 12 (y\partial_y - x \partial_x) K_{n}^{2a+1}(x,y).
\ee
This relation was first obtained for the Gaussian case in \cite{V2}.
Since it has been shown that $K_{n}^{2a+1}(x,y)$ 
is universal \cite{ADMN}, we thus have proved that the pre-kernel is
universal \cite{Senerprl}. 
The only problem is that an integral of the pre-kernel over the complete
spectrum
contributes to the spectral density and the spectral correlators.
Even in the microscopic limit this results in contributions 
from non-universal regions. Before going to the
general case, we first  analyze in detail the chGOE and the
case of a quartic probability potential.

\section{Two Examples}
In this section we study the quadratic
 and the  quartic probability potential.
In the first case, the  matrix
elements of $L$ and the skew-orthogonal polynomials  will be 
derived  exactly, whereas in the second case
only asymptotic results for large order polynomials will be obtained.

\subsection{The chiral Gaussian Orthogonal Ensemble}

In this section we study the chiral Gaussian Orthogonal Ensemble
by means of the operator construction discussed in the previous
section. The weight function is given by
$w(x) = x^{2a+1} e^{-nx}$, and the corresponding orthonormal
polynomials are the Laguerre polynomials,
\be
P_k^{2a+1}(x) =\frac 1{\sqrt {s_k^{2a+1}}} L_k^{2a+1}(nx),
\ee
with normalization constants
\be
s_k^{\alpha} = \frac {h_k^{\alpha}}{n^{\alpha + 1}} \quad {\rm and} \quad
h_k^\alpha = \frac {\Gamma(k+\alpha+1)}{k!}.
\ee 
The matrix elements of $L$ 
follow immediately from
the recursion relation
\be
x\partial_x L_k^{2a+1}(x) = kL_k^{2a+1}(x) -(k+2a+1)L^{2a+1}_{k-1}(x),
\ee
and are given by
\be
L_{kl} = \frac 12[(l+2a +1)\sqrt{ \frac {h_k^{2a+1}} {h_l^{2a+1}}}
 \delta_{k,l-1} -
(k+2a +1)\sqrt{ \frac {h_l^{2a+1}} {h_k^{2a+1}}} \delta_{l,k-1}].
\ee
One easily verifies that the recursion relation (\ref{lexplit}) does not
have a solution for  diagonal matrices $T_{kl}$. A solution is
obtained by  taking
$T_{2k,2k},\,\,\, T_{2k+1,2k+1},\,\,\, T_{2k+1,2k}$ and
$T_{2k+1,2k-1}$ as the only nonzero coefficients. In the
normalization  $R_{2k}(x)= L^{2a+1}_{2k}(nx)/\sqrt{s_{2k}^{2a+1}}$ 
(i.e. $T_{2k,2k} = 1$)
 the recursion relations (\ref{tjt}) simply read
\be
 T_{2k+1,2k+1} = L_{2k+1,2k},\qquad
 T_{2k+1,2k-1} = L_{2k,2k-1},
\ee
and the skew-orthogonal polynomials are thus given by
\be
R_{2k}^{a}(x) &=&\frac 1{\sqrt{s_{2k}^{2a+1}}}
  L^{2a+1}_{2k}(nx),\nonumber \\
R_{2k+1}^{a}(x) &=& -\frac{(2k+2a+2)}2 \frac{\sqrt{s_{2k}^{2a+1}}}
{s_{2k+1}^{2a+1}}
 L^{2a+1}_{2k+1}(nx)+ 
 \frac{(2k +1+ 2a)}2 \frac 1{\sqrt{s_{2k}^{2a+1}}}
L^{2a+1}_{2k-1}(nx)\nonumber \\
&+&T_{2k+1,2k} L_{2k}^{2a+1}(x).
\label{skewlaguerre}
\ee 
The coefficients $T_{2k+1, 2k}$ are not fixed by the orthogonality 
relations. Indeed, this is the well-known property that the
odd order skew-orthogonal polynomials are only determined up to a 
multiple of the even order polynomials of one degree lower. 
One can verify that these polynomials are normalized according
to $\langle R_{2k+1}^{2a+1}, R_{2k}^{2a+1} \rangle = 1$, and, with
an adjustment of the normalization, they coincide with the polynomials
obtained in \cite{V2} for a specific choice of
coefficient $T_{2k+1,2k}$.  

In fact, we can calculate the pre-kernel directly from
(\ref{pret}) using the matrix elements of $T^T JT$
without relying on explicit expressions for the skew-orthogonal
polynomials. Because only one term contributes to $R_{2k}(x)$, there 
are no terms in (\ref{lexplit})  that are outside the summation range
in (\ref{pret}). We thus have
\be
k_R(x,y)^a = \sum_{k,l = 0}^{2 \bar n-1} \frac 1{\sqrt{s_k^{2a+1} s_l^{2a+1}}}
L^{2a+1}_k(nx)L_{kl} L^{2a+1}_l(ny)
=\frac 12 (y\partial_y - x \partial_x) K^{2a+1}_{n}(x,y),
\ee
which was first obtained in \cite{V2} from the explicit properties of
the skew-orthogonal polynomials.
Using recursion relations for the Laguerre polynomials the pre-kernel
can be rewritten as
\be
\frac 12 (y\partial_y - x \partial_x) K^{\alpha}_{n}(x,y)
&=&\sum_{k=0}^{2 \bar n-1}\frac{(k+\alpha)}{2s_k^{\alpha}}
(L_{k-1}^{\alpha }(nx)L_{k}^{\alpha -1}(ny)-
L_{k-1}^{\alpha }(ny)L_{k}^{\alpha -1}(nx))
\nonumber \\
&=&\frac 12 (\partial_y -  \partial_x) K^{\alpha -1}_{n}(x,y),
\ee
where the  factor $k+\alpha$ 
has been absorbed in the normalization of the orthogonal 
polynomials.

According to (\ref{spect}), the  spectral density is given by
\be
\frac 12 \int_0^x dz e^{-\phi_a(z)-\phi_a(x)} 
(\partial_z -  \partial_x) K_{n}^{2a}(x,z)
-\frac 14
\int_0^\infty dz e^{-\phi_a(z)-\phi_a(x)}  
(\partial_z -  \partial_x) K_{n}^{2a}(x,z)
\ee
The microscopic limit of the first term 
results in an integral over the universal Bessel
kernel. However, it is not possible to 
the interchange the integral and the microscopic
limit in the second term. To see this we return to the definition of
the pre-kernel. Then the second term is given by 
\be
-\frac 14
\int_0^\infty dz  e^{-\phi_a(z)} \sum_{i,j=0}^{2\bar n - 1}
R_i^{a}(x) J_{ij} R_j^{a} (z) e^{-\phi_a(x)}.
\ee
We thus consider the integral
\be
\int_0^\infty dz e^{-\phi_a(z)} R_i^{a}(z) = 
\int_0^\infty dz z^ae^{-nz/2} R_i^{a}(z).
\ee
By using the explicit expressions for the skew-orthogonal polynomials
and the relation
\be
L^{2a+1}_n(2x) = \sum_{m=0}^nL^{a}_{n-m}(x) L^{a}_m(x),
\ee
it follows that the integral over the odd order 
skew-orthogonal polynomials vanishes. 
The integral over the even order skew-orthogonal polynomials is up
to a normalization constant given by
\be
 \frac{(n/2)^{a+1} }{h_{p}^a} \int_0^\infty dz z^ae^{-nz/2} 
L_{2p}^{2a+1}(nz) = 1.
\ee
Let us now calculate the integral by interchanging the integration and the
microscopic limit. 
Using the asymptotic relation relation between Laguerre polynomials and Bessel functions
\be
L_n^\alpha(nx) \sim x ^{-\alpha/2} J_\alpha(2 n\sqrt{x})
\label{laguerreasymp}
\ee
the microscopic limit of the integral is given by
\be
\int_0^\infty   \lim_{\tiny \begin{array}{c} p\rightarrow \infty \\
p/n=fixed \end{array}} dz
z^ae^{-nz/2 }  \frac{(n/2)^{a+1} }{h_{p}^a}
L_{2p}^{2a+1}(nz) = \int_0^\infty dw J_{2a+1}(2w)= 
\frac 12 .
\ee
Exactly half of the integral is missing. In the next section we will
argue that this is a general feature of the skew-orthogonal polynomials.
However, the integral over the odd-order polynomials does not vanish in
general. The  result for the microscopic spectral density is
thus given by
\be
\rho_s(u) = \frac 12 \int_0^u dw   (\partial_w -  \partial_u) 
{B}^{2a}(u,w)
-\frac 12 \int_0^\infty dw  (\partial_w -  \partial_u) B^{2a}(u,w),
\ee
where an additional factor of 2 has been included in the second term and
the Bessel kernel is defined in (\ref{besselkern}).

\subsection{Chiral Orthogonal Ensemble with Quartic Potential}

In order to construct
 the skew-orthogonal polynomials using the Br\'ezin-Neuberger
formalism \cite{BN}, we need an expression for the derivative of orthogonal
polynomials.  To derive such relation
we start from the recursion relation \cite{ADMN}
\be
x P_k(x)  = -r_k(P_{k+1} -P_k) + s_k(P_k -P_{k-1}) .
\label{recur}.
\ee
The coefficients $r_k$ and  $s_k$ are related by
\be
s_k =\frac{h_k r_{k-1}}{h_{k-1}},
\ee
and $h_k =\int_0^\infty dx w(x) P_k^2(x)$ is the normalization integral.
The recursion relation (\ref{recur}) is valid for orthogonal 
polynomials normalized according to $P_k(0) = 1$. To make contact with the
analysis of \cite{ADMN} we will use this normalization in this section.

The derivative of the polynomials for arbitrary weight function
is given by \cite{Kanz97}
\be
y P_k'(y) &=& \int_0^\infty dx w(x) 
\sum_{l=0}^k \frac{P_l(x)P_l(y)}{h_l} x P_k'(x)
\nonumber \\
&=&kP_k(y) +\int_0^\infty dx w(x)
 \sum_{l= 0}^{k-1} \frac{P_l(x)P_l(y)}{h_l} x P_k'(x)\nonumber \\
&=& kP_k(y) -\int_0^\infty dx w'(x) 
\sum_{l=0}^{k-1}\frac{P_{l}(x)P_{l}(y)}{h_{l}} x P_k(x).
\label{derivative}
\ee
where the terms following the last equal sign have been obtained 
by partial integration.

Next we derive the asymptotic form of large order 
skew-orthogonal polynomials for a quartic probability potential. In terms
of the $x_k=\lambda^2_k$ the weight function is given by  
$w(x) = x^{2a+1}e^{-nx^2/2}$. Only the terms  $l = k-1$ and
$l = k-2$ are nonvanishing in the last sum in (\ref{derivative}). Using
 the recursion relation (\ref{recur}) to calculate the integrals we 
find 
\be
y P_k'(y)=
 kP_k(y) 
-n s_k[r_k+s_k +r_{k-1} +s_{k-1}]P_{k-1}(y)
+ n s_k s_{k-1} P_{k-2}(y).
\ee
Taking into account the normalization of the orthogonal polynomials
the matrix elements of $L$ are given by
\be
L_{kl} &=& \frac 12\int_0^\infty dx w(x) 
\frac 1{\sqrt{h_k h_l}}[P_l(x) xP'_k(x)-P_k(x) xP'_l(x)]
\nonumber \\
&=&\delta_{l,k-2}\frac n2 \sqrt{\frac {h_{k-2}}{h_k}}s_ks_{k-1}
-\delta_{l,k-1}\frac n2 \sqrt{\frac {h_{k-1}}{h_k}}
 s_k[r_k+s_k +r_{k-1} +s_{k-1}]\nonumber \\
&+&\delta_{k,l-1}\frac n2 \sqrt{\frac {h_{l-1}}{h_l}}
 s_l[r_l+s_l +r_{l-1} +s_{l-1}]
-\delta_{k,l-2} \frac n2 \sqrt{\frac {h_{l-2}}{h_l}}s_ls_{l-1}.\nonumber \\
\ee
As expected, the  $L_{kl}$ vanish for $|k-l| > 2$.

In the limit $n \rightarrow \infty$ the leading order contributions
to the kernel are for terms with large values of $k$ and $l$.
The large-$k$ asymptotic behavior of the
coefficients in the recursion relation (\ref{recur}) 
was obtained in \cite{ADMN}
for an arbitrary polynomial probability potential (but for integer 
values of $2a$).
For fixed $t\equiv k/n$, 
the continuum limit of the coefficients can be parameterized as
\be
h_k = \frac 1{n^{4a+3}} h(t), 
\qquad s_k = s(t), \quad {\rm and} \quad r_k = r(t).
\ee
For $k \to \infty$, only $s_k $ and $r_k$ enter in the matrix elements of
$L$. They are given by \cite{ADMN}
\be
r(t) = s(t) = \sqrt{\frac t3}.
\ee
The leading order asymptotic result for the matrix elements of $L$ thus
reads
\be
L_{kl} 
=\delta_{l,k-2} \frac {nt}6 
-\delta_{l,k-1} \frac {4nt}6
+\delta_{k,l-1} \frac {4nt}6
-\delta_{k,l-2} \frac {nt}6.
\ee
By inspection of the recursion relation (\ref{tjt}) one easily finds 
that for a quartic potential (i.e. $p=2$) 
the skew-orthogonal polynomials  can be expressed  as
\be
R_{2k} (x)  &=& \tilde P_{2k}(x) + T_{2k,2k-1} \tilde P_{2k-1}(x),\\
R_{2k+1} (x)  &=& T_{2k+1,2k+1}\tilde P_{2k+1}(x) + 
T_{2k+1,2k} \tilde P_{2k}(x)
+ T_{2k+1,2k-1} \tilde P_{2k-1}(x)+ T_{2k+1,2k-2} \tilde 
P_{2k-2}(x),\nonumber \\
\ee
where the orthonormal polynomials have been denoted by
$\tilde P_k(x) \equiv P_k(x)/\sqrt h_k$.
The coefficients $T_{i,k}$ obey the recursion relations
\be
T_{2k-1,2k-1} - T_{2k,2k-1}T_{2k+1,2k-2}= L_{2k-1,2k-2}, & 
\quad & T_{2k+1,2k-2}  =- L_{2k,2k-2},
\nonumber \\
T_{2k,2k-1}  T_{2k+1,2k}-T_{2k+1,2k-1} =  L_{2k,2k-1}, & \quad & 
T_{2k,2k-1}T_{2k+1,2k+1}  =  L_{2k+1,2k-1}.
\label{tall}
\ee
Using the asymptotic values for the matrix elements of $L$ we
find a recursive equation
for the coefficients $T_{2p,2p-1}$, 
\be
T_{2k,2k-1} T_{2k-2,2k-3}+4T_{2k-2,2k-3} +1=0.
\label{tk}
\ee
 This recursion relation has two fixed points given by
the roots of
\be
x^2 +4x + 1= 0.
\ee
For large $k$ the coefficients $T_{2k,2k-1}$ should depend  smoothly on   $k$
and are thus given by one of the fixed points. The polynomial corresponding
to the stable fixed point,
 $x= -2 -\sqrt 3$, given by  
$\tilde P_{2k}(x) -(2+\sqrt 3) \tilde P_{2k-1}(x)$, is negative for $x = 0$
(we work in the convention $P_l(0) = 1$, and 
$ \tilde P_{2k}(0)$ and $\tilde P_{2k-1}(0)$ are equal to leading
order in $1/k$)   and positive for 
$x\rightarrow \infty$ and thus has an odd number of zeros. 
Since even orthogonal polynomials should have an even number
of zeros, the 
relevant solution is thus given by the unstable fixed point
$x = -2 +\sqrt 3$. This counter-intuitive result is a reflection of the
numerical instability of the iterative 
construction of orthogonal polynomials. A numerical confirmation will
be given in section 7.
All other coefficients simply follow
from the  relations (\ref{tall}) resulting in the polynomials
\be
R_{2k} (x)  &=& \tilde P_{2k}(x) -(2-\sqrt 3) \tilde P_{2k-1}(x),\nonumber \\
R_{2k+1} (x)  &=& \frac {-nt}{6(2-\sqrt 3)}\tilde P_{2k+1}(x) + 
 \frac {4nt}{6} \tilde P_{2k-1}(x) -\frac {nt}6 \tilde P_{2k-2}(x)
\nonumber \\ &+& T_{2k+1,2k}( \tilde P_{2k}(x) -(2-\sqrt 3)\tilde  
P_{2k-1}(x)).
\label{quarticskew}
\ee
However, we do not need the explicit expressions for the
skew-orthogonal polynomials. The pre-kernel can be 
expressed directly in the matrix elements of $L$ and two additional terms
that are outside the summation range in
(\ref{pret}),
\be
k_R(x,y) &=& \sum_{k,l = 0}^{2\bar n-1} \frac 1{\sqrt{h_k h_l}}
P_k(x)L_{kl} P_l(y)\nonumber \\
&+&\frac {T_{2\bar n,2\bar n-1}T_{2\bar n+1,2\bar n-2}}
{\sqrt{h_{2\bar n-1} h_{2\bar n-2}}}
[P_{2\bar n-1}(x)P_{2\bar n-2}(y)- P_{2\bar n-1}(y)P_{2\bar n-2}(x)].
\nonumber \\
\label{krsum}
\ee
Because $\lim_{n\rightarrow \infty} T_{2n,2n-1} = -2 +\sqrt 3$, the
additional terms are of the same order of magnitude as each of
the terms in the sum. Therefore, to leading order in $1/n$ 
the relation between the pre-kernel for $\beta =1$ and the kernel for
$\beta = 2$ is the same as for the Gaussian case. 
However, it is not clear whether the interchange of the integral 
and the microscopic limit in $K_R(\infty,x)$  also proceeds in the
same way. This question will be analyzed
in the next section for an arbitrary probability potential.

\section{A property of large order skew-orthogonal polynomials}
\label{property}
In this section we will derive an  asymptotic relation   for
 large order skew-orthogonal
polynomials. Our starting point is the skew-orthogonality relation 
\be
\langle R_k, R_0 \rangle_R =
\int_0^\infty dx\int_0^\infty dy
R_k(x) e^{-\phi_a(x)-\phi_a(y)} \epsilon(x-y)  = 0.
\ee
For asymptotically large $k$ we can distinguish three different domains
in the integration over $x$. 
The region near $x=0$, the
region around the largest zero of $R_k(x)$, and the oscillatory
region which is in between these two regions. We split the integration    
over
$x$ into two parts separated by $M$ chosen to be inside the oscillatory    
region. In our normalization the spacing of the smallest eigenvalues
scales as $\Delta x\sim 1/n^2$. 
The main contribution to the integral over $[0,M]$ is from the region    
around the smallest eigenvalues
$x \sim 1/n^2$, whereas the $y$-integral has contributions up to 
$nV(y) \sim 1$. If the potential behaves as $y^p$ with $p>1/2$ near $y =    
0$,
to leading order in $1/n$ the contribution to the $y-$integral is from 
the region with $y>x$ and $\epsilon(x-y) = -1/2$. The main contribution
to the integral over $[M,\infty \rangle$ is from the region near the
largest zero of $R_k(x)$. To leading order in $1/n$ we can then
replace $\epsilon(x-y) \rightarrow 1/2$. The integrals over $x$ and $y$
factorize and we obtain the following asymptotic relation
\be
\int_0^M dx R_k^a(x) e^{-\phi_a(x)} dx = \int_M^\infty dx R_k^a(x)
 e^{-\phi_a(x)}    
dx .
\label{intM}
\ee
This implies that 
\be
\int_0^\infty dx R_k^a(x) e^{-\phi_a(x)} dx = 
2\int_0^M dx R_k^a(x) e^{-\phi_a(x)} dx,
\label{preas}
\ee
which is valid for $k\rightarrow \infty$ provided that the r.h.s. is
independent of $M$ in the oscillatory region. 

In our derivation we have made the assumption that
the skew-orthogonal polynomials show
the same oscillatory behavior as the regular orthogonal polynomials.  
This is certainly true 
if they can be expressed in a finite number of regular orthogonal    
polynomials which is the case for a finite order polynomial
probability potential \cite{Senerprl}. Of course, a different 
oscillatory behavior is possible if the leading order 
asymptotic terms in the skew-orthogonal polynomials cancel. 
This results in contributions
that are subleading in $1/n$. A  priori, it cannot be excluded that
the skew-orthogonal polynomials near zero and the l.h.s. of (\ref{preas}) are
of the same order in $1/n$ and thus both contribute to the kernel.  
For example, this is the case for
the odd-order skew-Laguerre polynomials (\ref{skewlaguerre})
with  $T_{2k+1,2k} = 0$. However, in this case the l.h.s. of
(\ref{preas}) vanishes, and we do not have to worry
about the asymptotic behavior of the r.h.s. of (\ref{preas}).
In general, there is no reason to expect that the leading order 
asymptotic expansion of the $P_k^{2a+1}$ cancels. For example, 
the asymptotic behavior of  $P_k^{2a+1}$ in the oscillatory
region depends smoothly on $k$, and from the explicit expressions
for the even order skew-orthogonal polynomials in
(\ref{quarticskew}) it then follows that the leading order 
asymptotic behavior does not cancel. For the odd order
skew-orthogonal polynomials a cancellation can only be achieved by 
fine tuning the coefficient $T_{2k+1,2k}$.

For a finite order probability potential, 
the generic asymptotic behavior of the  $R_k^a(x)$ is thus
the same as that of  $e^{-\phi_a(x)}P_k^{2a+1}(x)$. In the region near
zero and in the oscillatory region,
it is given by $J_{2a+1}(c\sqrt{k x})/\sqrt x$ (with $c$ a constant that can
be obtained from the recursion relations). 
If the integral (\ref{preas})
is nonvanishing to leading order in $1/n$, we thus find from the 
asymptotic behavior of  $e^{-\phi_a(x)}P_k^{2a+1}(x)$
that the integral converges for $k \rightarrow \infty$ and
 $M$ inside the oscillatory region.

As will be shown below, the even order skew-orthogonal polynomials
are determined up to a multiplicative constant.  
{}From the orthogonality
relations it is clear that the odd order polynomials $R_{2k+1}(x)$
are only determined up to the addition of a multiple of $R_{2k}(x)$. 
We may use this freedom to chose a normalization such  that 
\be
\int_0^\infty dx e^{-\phi_a(x)} R_{2k+1}^a(x) = 0.
\label{assymodd}
\ee
In this way we avoid the ambiguity in the asymptotic behavior of the
odd order skew-orthogonal polynomials.

The  main contributions to the integral in the r.h.s. of (\ref{preas}) 
are from
the region close to $x = 0$ and the integrand can be replaced by its
microscopic limit. Our final result is
\be
\lim_{\tiny \begin{array}{c} n\rightarrow \infty \\ k/n=const. \end{array}}
\int_0^\infty R_{2k}^a(x) e^{-\phi_a(x)} dx = 2\int_0^\infty 
\lim_{\tiny \begin{array}{c} n\rightarrow \infty \\  x n^2 =z^2 
\\ k/n=const. \end{array}} R_{2k}^a(x) e^{-\phi_a(x)}    
dx. \nonumber \\
\label{aseven}
\ee
Interchanging the microscopic limit and the integral gives rise to
an extra factor
two. If the integrand in the kernel $K_R^a(\infty,x)$ is replaced by
its microscopic limit, the same extra factor two has to be included,
 \be
\lim_{\tiny \begin{array}{c}n\rightarrow \infty  \\  xn^2 = z \end{array}}
K_R^a(\infty,x) = 2 \int_0^\infty 
 \lim_{\tiny \begin{array}{c} n\rightarrow \infty \\x n^2 =z\end{array}} 
dw e^{-\phi_a(w/n^2) - \phi_a(z/n^2) } \frac 1{n^2} 
k_R^a(\frac{w}{n^2},\frac{z}{n^2}).
\ee 
We emphasize that this relation is based on the asymptotic
properties of the even order skew-orthogonal polynomials only.

\section{Universality of the Chiral Orthogonal Ensemble}

\label{general}

The proof of universality of microscopic spectral correlation
functions for the chiral orthogonal ensemble for a finite polynomial
probability potential is now straightforward.
It is an immediate consequence of the following three results. i) The relation
between the pre-kernel for the chOE and the chUE kernel is independent
of the probability potential to leading order in $1/n$ \cite{Senerprl}. 
ii) The 
microscopic limit of the chUE kernel is universal if the eigenvalues
are expressed in units of the average level spacing \cite{ADMN}. 
iii) Because of a novel
asymptotic property of large order skew-orthogonal polynomials, the
microscopic limit and the integrals that occur in the spectral correlation
functions can be interchanged at the expense of a factor of 2.

All $k$-point correlation functions of the chOE can  be expressed compactly
as quaternion determinants of quaternions \cite{mm}
\be
S(x_k,x_l) = \mat K_{R,n}^a(x_k,x_l) - \frac 12 K_{R,n}^a( \infty,x_l) &
 \int_{x_l}^{x_k} dz [K_{R,n}^a(x_k,z)-
\frac 12 K_{R,n}^a(\infty,z)] -\epsilon(x-y)\\
\partial_{x_k} K_{R,n}^a(x_k,x_l) &  K_{R,n}^a(x_l,x_k) - 
\frac 12 K_{R,n}^a( \infty,x_k) \emat \nonumber \\
\ee
where $K_R^a(x,y)$ is defined in (\ref{KR}).
We already noticed that the spectral density is given by 
$\rho(x) = K_{R,n}^a(x,x) - \frac 12 K_{R,n}^a( \infty,x)$. 
The two-point cluster function is given by \cite{mm}
\be
T(x,y) = \frac 12 {\rm Tr} S(x,y) S(y,x).
\label{txy}
\ee
The universal microscopic kernel for the chOE is obtained
by taking the microscopic 
limit of the integrands and replacing the factors $1/2 \to 1$,
\be
S(u,v) &=& \lim_{N=2n\to \infty} \frac {2\sqrt{uv} }{\Sigma^2 N^2} 
S(\frac{u^2}{\Sigma^2 N^2,}\frac{v^2}{\Sigma^2 N^2})\nonumber \\
&=& 
2\sqrt{uv} \mat Q^a(u,v) - Q^a(\infty,v) & 
{\Sigma^2 N^2}\int_{v^2}^{u^2} dw^2 [Q^a(u,w) - Q^a(\infty,w)]
\\ \frac 1{\Sigma^2 N^2} \partial_{u^2} Q^a(u,v) & 
Q^a(v,u) - Q^a(\infty,u)\emat,\nonumber \\
\ee
where $Q^a(u,v)$ is the microscopic limit of the kernel $K^{a}_{R,n}(x,y)$.
Universality then follows from the relation between the microscopic
kernels for $\beta=1$ and $\beta =2$ and the universality of the
microscopic limit of the kernel for $\beta =2 $.
The universal microscopic result for $Q^{a}(u,v)$ is given by
\be
Q^{a}(u,v) &\equiv&	
\lim_{N=2n\to \infty} \frac {1}{\Sigma^2 N^2}
 K_{R,n}^{2a}(
\frac{u^2}{\Sigma^2 N^2},\frac{v^2}{\Sigma^2 N^2})
\nonumber \\ &=&
\frac 14 \int_0^{u^2} d[w^2] (wv)^{2a}  (\partial_{w^2} -
\partial_{v^2}) (wv)^{-2a-1/2} B^{2a}(v,w).
\ee
By using the integral representation (\ref{besselkernint}) of
the Bessel kernel we find the following explicit representation
of the kernel
\be
Q^{a}(u,v) = \frac u4 \int_0^1 dw 
\int_0^1 t^2 dt 
\left [\frac {uw }{v} J_{2a}(uwt) J_{2a+1}(vt)
-J_{2a+1}(uwt) J_{2a} (vt) \right ].
\ee
One of the integrals can be performed analytically. By using identities for
Bessel functions one ultimately finds the result
\be
2\sqrt{uv} Q^{a}(u,v) &=& B^{2a+1}(u,v) - \sqrt{uv} \frac{J_{2a+1}(v)}{2v} 
\left ( \int_0^u dw J_{2a+1}(w) -1\right ) \nonumber \\
   &=& B^{2a}(u,v) - \sqrt{uv} \frac{J_{2a+1}(v)}{2v} 
\left ( \int_0^u dw J_{2a-1}(w) -1\right ).
 \ee
The microscopic spectral density for the chOE is 
 given  by $\rho_s(u)= 2u(Q^a(u,u) - Q^a(\infty,u))$. For its universal
form we thus obtain \cite{V2}
\be
\rho_s(u) &=& \frac {u^2}2 \int_0^1 t^2dt\int_0^1 dw 
\left [ w J_{2a}(uwt)
J_{2a+1}(ut) - J_{2a+1}(uwt) J_{2a}(ut)\right ]
+\frac 12 J_{2a+1}(u).\nonumber \\
\ee
This expression for the microscopic spectral density can be simplified to
\be
\rho_s(u)
&=& \frac u2\left [J_{2a+1}^2(u) - J_{2a+2}(u) J_{2a}(u) \right ]
- \frac 12
J_{2a+1}(u) \left ( \int_0^u dw J_{2a+1}(w) -1 \right )\nonumber \\
&=& \frac u2\left [J_{2a}^2(u) - J_{2a+1}(u) J_{2a-1}(u) \right ]
- \frac 12 J_{2a+1}(u) \left ( \int_0^u dw J_{2a-1}(w) -1 \right ).
\ee
The simplified results for $\rho_s(u)$
and the kernel were first obtained in \cite{Tilohab} and was derived
independently in \cite{forrester14} (with a typo, 
see also \cite{Widom,Nishim,akekan}). 
The first term can be 
recognized as the microscopic spectral density for the chGUE
\cite{VZ,Vinst}.
The microscopic limit of the two-point 
correlation function follows immediately
from (\ref{txy}) and the microscopic limit of the kernel.
 
Recently, universal results for massive spectral correlators at
$\beta = 1$ and $\beta = 4$
have been obtained by relating the kernels for the massive correlators
to the corresponding massless kernel \cite{akekan}. 
Similar relations have been derived for the Gaussian case \cite{Nishim}.

\section{Numerical study of the $x^4$-potential}
\label{numerical}
In this section we  explicitly construct the skew-orthogonal polynomials for
an $x^4$ potential and test the  asymptotic results obtained in previous
sections. We consider the distribution    
of the squared eigenvalues $x_k = \lambda_k^2$ on  $[0, \infty \langle$ with  
 weight  function $\phi_a(x) = x^2/2 -a\log x$. For reasons of numerical
accuracy we only consider integer values of $a$.  

\begin{figure}[ht!]
     \centerline{ \epsfig{file=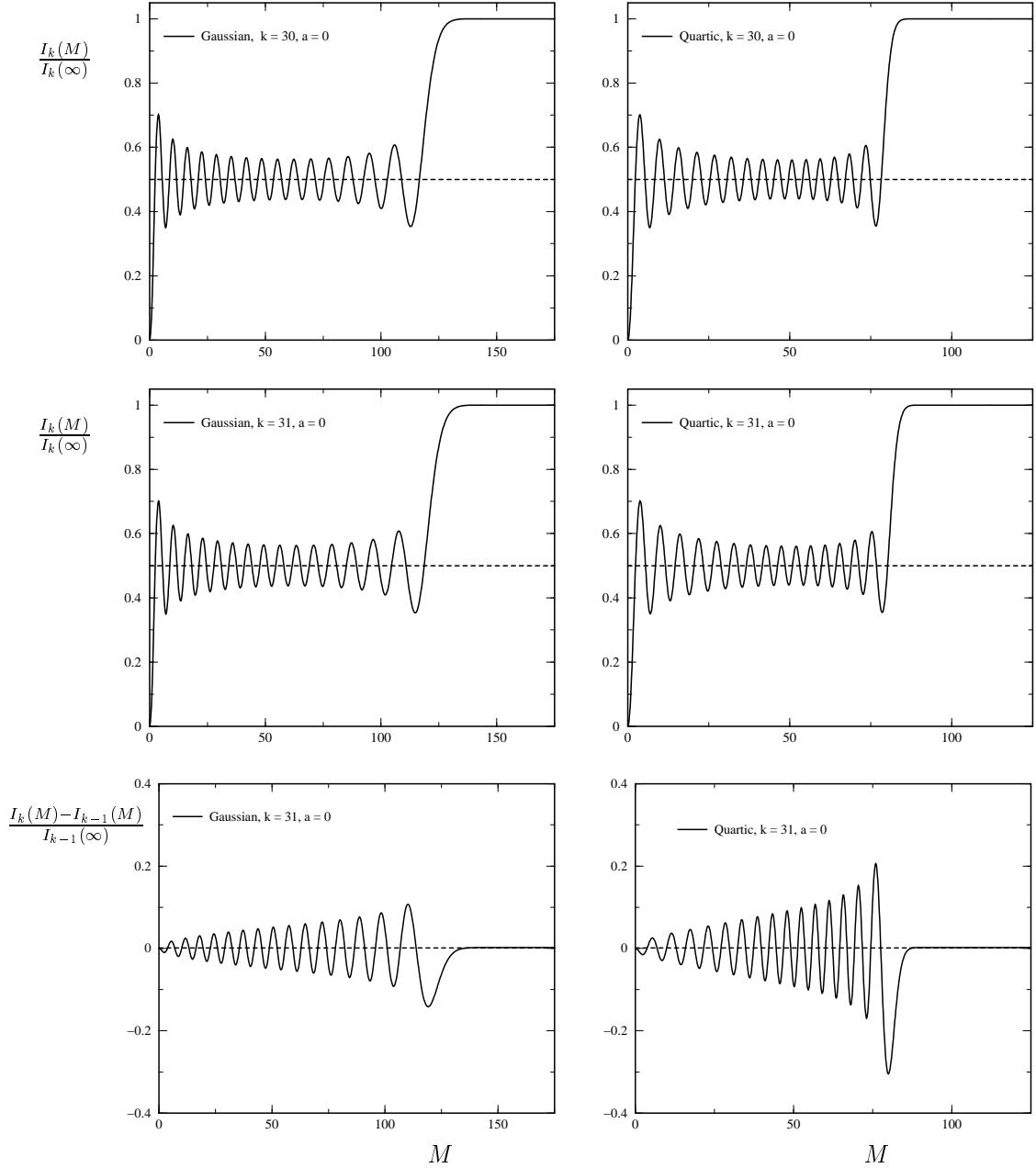,width=18.5cm}}
\vspace*{-4cm}
\caption{\label{} Ratios of the integral 
$I_k(M)=\int_0^{(M/\pi\rho(0))^2} dx \, e^{-\phi_a(x)} R_k(x)$ versus $M$
 for a   Gaussian (left) and a 
quartic (right) probability  potential. 
The  parameter $a$ and 
the value of $k$ are given in the label of the figures.}
\end{figure}    

The skew-orthogonal polynomials can be expanded in terms of monomials    
$x^k$ as 
\be
R^{a}_k(x)= r^{a}_{kk} x^k + r^{a}_{k, k-1} x^{k-1} + \cdots +    
r^{a}_{k0}.
\ee
The coefficients are determined from the orthogonality relations
(\ref{orthoskew}). 
They can be reduced to a system of linear equations. For the    
polynomials of even order, $R^{(a)}_{2k}$, one finds
\be
\begin{array}{*{3}{c@{\; + \;}}c@{\;=\;}c}
t^{a}_{00}r^{a}_{2k,0} & t^{a}_{01}r^{a}_{2k,1} & \cdots &    
t^{a}_{0,2k}r^{a}_{2k,2k} & 0, \\
t^{a}_{10}r^{a}_{2k,0} & t^{a}_{11}r^{a}_{2k,1} & \cdots &    
t^{a}_{1,2k}r^{a}_{2k,2k} & 0, \\
t^{a}_{20}r^{a}_{2k,0} & t^{a}_{21}r^{a}_{2k,1} & \cdots &    
t^{a}_{2,2k}r^{a}_{2k,2k} & 0, \\
\multicolumn{5}{c}{\dotfill}\\
t^{a}_{2k-1,0}r^{a}_{2k,0} & t^{a}_{2k-1,1}r^{a}_{2k,1} & \cdots &    
t^{a}_{2k-1,2k}r^{a}_{2k,2k} & 0.
\end{array} 
\label{homo}
\ee
For the polynomials of odd order $2k+1$, the first $2k-1$ 
equations have one more term,  $t^{a}_{i,2k+1}r^{a}_{2k+1,2k+1}$, and   
the  $r^{a}_{2k,i}$ are replaced by $r^{a}_{2k+1,i}$. The normalization
equation 
$\langle R_{2k+1}, R_{2k} \rangle_R= 1$ reads
\be
r^{a}_{2k,2k}[
t^{a}_{2k,0}r^{a}_{2k+1,0} + t^{a}_{2k,1}r^{a}_{2k+1,1} +\cdots +    
t^{a}_{2k,2k+1}r^{a}_{2k+1,2k+1}] = -1.
\ee

The $t^{a}_{i,j}=\langle x^i, x^j \rangle_R$ are the skew-scalar 
products of    
the monomials. By partial integration, it is possible to derive a    
recursion relation relating $t_{i,j}^a$ and $t_{i, j-2}^a$,
\be
t_{k,l}^a &=& \langle x^k, x^l \rangle_R\nonumber \\
        &=& \int_0^\infty dx x^kx^a e^{-x^2/2}\int_0^\infty dy y^l
y^a e^{-y^2/2}
            \epsilon(x-y)\nonumber \\ 
        &=& (l+a-1)t_{k,l-2}^a - 
\frac 12 \Gamma\left ( \frac {k+l+2a}2 \right ).
\label{recurt}
\ee
Using the antisymmetry    
of $t_{i,j}^a=-t_{j,i}^a$, 
all skew-scalar products can  then be easily calculated from    
$t_{0,0}^a=t_{1,1}^a=0$ and $t_{0,1}^a$. For a weight function with    
positive integer parameter $a$, the skew-scalar products 
are related to the case    
$a=0$ by $t^a_{i,j}=t^0_{i+a, j+a}$. 
Because of the antisymmetry of the skew-scalar product, 
$\langle R^{a}_i, R^{a}_i \rangle=0$, but  this relation    
does not impose an additional condition on the coefficients.

We  construct the skew-orthogonal polynomials from the homogeneous
equations (\ref{homo}). They can be easily normalized later by multiplying the
even or odd order polynomials by a suitable scale factor.
For  the coefficients of the even order polynomials, the number of
equations is one  less than the number of coefficients, 
whereas  for the odd order ones
we lack two equations. To determine the polynomials, we fix
 $r^{a}_{2k,2k} = 1$, $r^{a}_{2k+1,2k+1} = 1$ and
$r^{a}_{2k+1,2k} = 0 $ (the latter condition can be imposed because
$R_{2k+1}$ is determined only up to a multiple of $R_{2k}$). 
In this way the polynomials can be determined accurately to about
order 30. The skew-orthogonal polynomials for the Gaussian case
can be derived in a similar way.

\begin{figure}[ht!]
\centerline{\epsfig{file=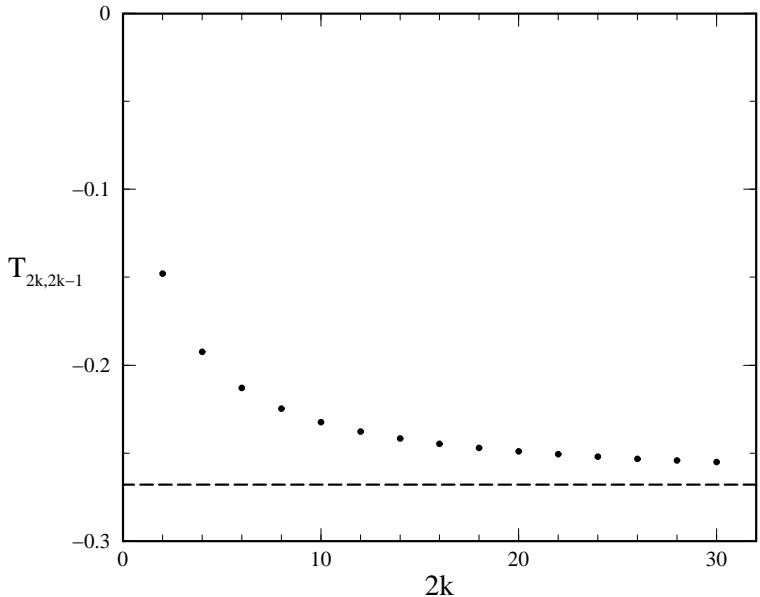,width=8.2cm,angle=270}}
 \caption{The asymptotic behavior of the coefficients $T_{2k,2k-1}$
versus $k$ for a quartic probability potential. Its asymptotic
value of $-2+\sqrt 3$ is depicted by the dashed curve.}
\label{rho1}
\end{figure}

To illustrate the asymptotic behavior of the skew-orthogonal polynomials
we show in Fig. 1 the $M$-dependence of the ratio $I_k(M)/I_k(\infty)$ 
(full curves) for a Gaussian (left) and a quartic (right) probability
potential both with parameter $a=0$ . The 
integral $I_k(M) $ is  defined by
\be
I_k(M) =\int_0^{(M/\pi\rho(0))^2} dx \, e^{-\phi_a(x)} R_k(x). 
\label{target}
\ee 
The weight function is given by
$\phi_a(x) = x/2$ for the Gaussian case and by 
$\phi_a(x) = x^2/2$ for the quartic case.
We have  redefined $M$ in units of
 $\pi \rho(0)$, with $\rho(0)$  the 
the average spectral
density near zero (notice that with our convention for
the weight function $\rho(0)$ has a nontrivial $k$-dependence).
 This figure shows that in
the intermediate domain the integral $I_k(M)$ oscillates
 around $I_k(\infty)/2$. 
For even $k$, the integral appears to converge 
in the
oscillatory region. For odd $k$ we show 
results for monic polynomials with normalizations $r_{2k+1,2k} = 0$ (middle
figures) and by 
$R_{2k+1}(0) = 0$ (lower figures). 
In the first case, the odd order polynomials
behave similarly to the even order ones, whereas in the
second case the behavior is quite different, because
the leading order asymptotic contributions cancel. 
The oscillations in the lower figures 
are still exactly about $I_k(\infty)/2$ which in this case is close
to zero. 
The even order skew-orthogonal polynomials always have as
many  positive zeros as the order of the polynomial. 
The odd order polynomials in the  normalization $r_{2k+1,2k} = 0$
are not very different from the preceding even order polynomials (see middle
figure). However, typically one of their zeros is located on the negative
real axis. For the normalization $R_{2k+1}(0) = 0$,
the total number of zeros is equal to $2k+1$,
with one zero at $x =0$.
Fortunately, 
as we have seen in the previous section, the
integral (\ref{assymodd}) over the odd order polynomials can always be
tuned to zero so that we do not have to worry about the ambiguity
of the asymptotic properties of the odd order skew-orthogonal polynomials.
 
\begin{figure}[ht!]
\setlength{\unitlength}{2.0in} 
\centerline{\epsfig{file=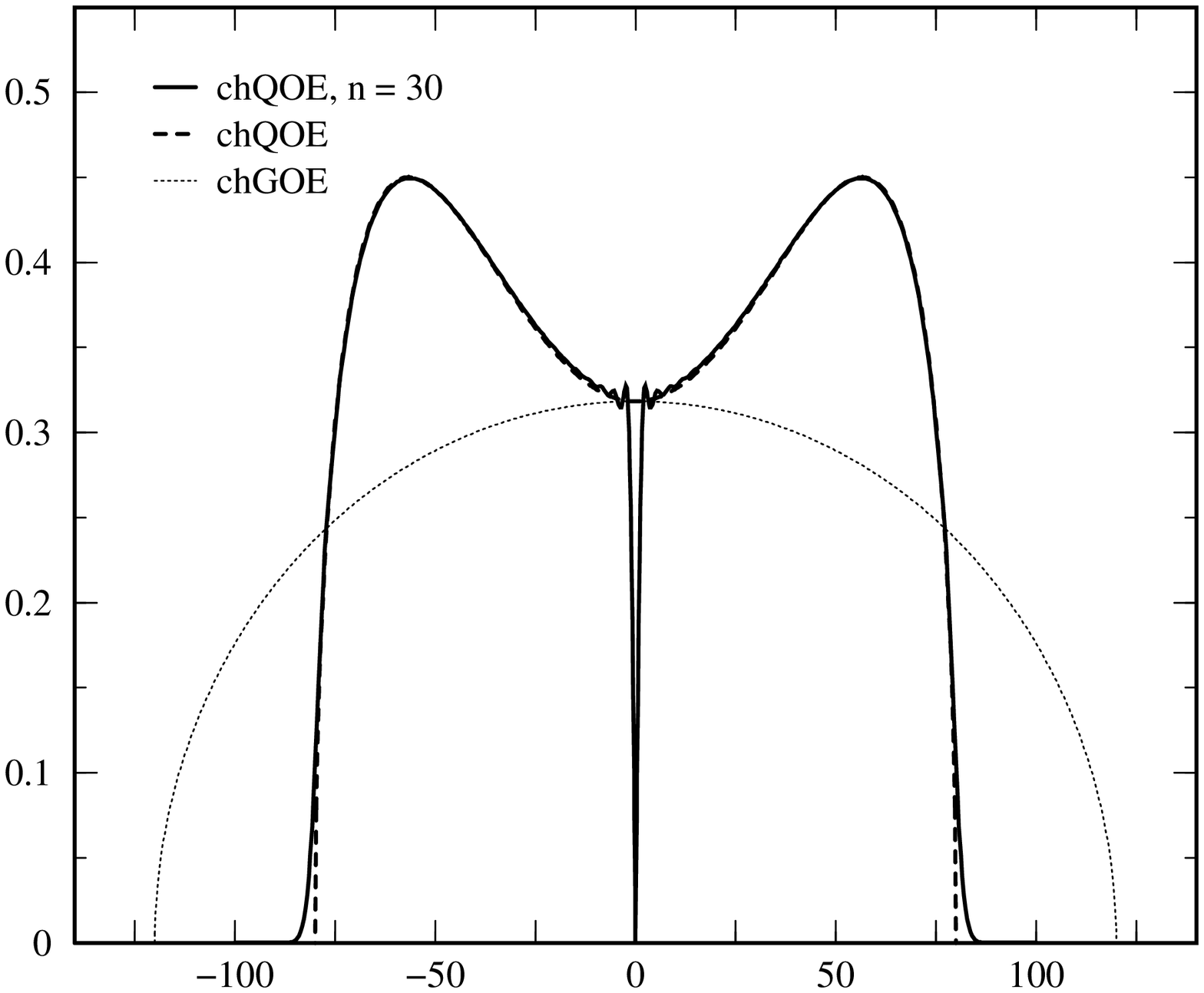,width=8.2cm,angle=270}}
\begin{picture}(1,1)                                                  
\put(1.6,0.9){\Large $u$}                                            
\put(0.2, 2.00){\Large $\frac{\rho \left ( 
\frac u{\pi\rho(0)} \right )}{\pi\rho(0)} $ }                                
\end{picture}   
\vspace*{-4cm}
 \caption{The average spectral density $\rho(u/\pi\rho(0))/\pi\rho(0) $ 
versus $u$. For a quartic probability potential
 we show results obtained
from the 30 lowest order skew-orthogonal polynomials (full curve) and the
large-$n$ analytical result (dashed curve). The semicircular
distribution (dotted curve) represents the
large-$n$ average spectral density for a Gaussian probability potential.
All results are for $ a= 0$.}
\label{rho}
\end{figure}

In Figure 2, we show the $k$-dependence of the 
coefficients $T_{2k,2k-1}$ as defined in equation
(\ref{quarticskew}). Also shown is the analytical result for the
asymptotic value of $-2+\sqrt 3$. The convergence to the
asymptotic result is better illustrated by
extrapolating to second order in $1/k$. The values of
 $T_{2k,2k-1}$ at $k=10,\,\,20$ and 30 extrapolate to
$-0.26787$ to be compared to $-2+\sqrt 3 = -0.26795$.

\begin{figure}[ht!]
\centerline{\epsfig{file=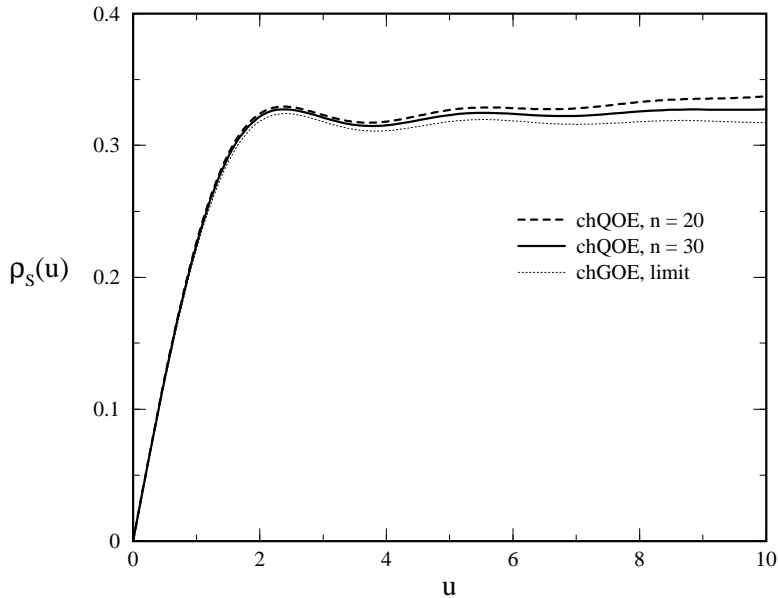,width=8.2cm,angle=270}}
 \caption{Microscopic spectral density $\rho_s(u)$ 
for a Gaussian and    
a quartic probability potential.}
\label{rho4}
\end{figure}

In Figure 3, we show the average spectral density calculated from
the quartic skew-orthogonal polynomials (full curve) and the analytical result
(dashed curve) given by 
\be
\frac{\rho(u/\pi\rho(0))}{\pi \rho(0)}=
\frac 1\pi \left ( 1 +\frac 3{4n} u^2 \right ) \left (
1 - \frac 3{8n} u^2 \right )^{1/2}.
\ee 
For comparison we also show the semicircular distribution
obtained for a Gaussian potential (dotted curve).

In Fig. 4 we show the microscopic spectral density calculated
from the first 30 skew-orthogonal polynomials for a quartic potential
and $a = 0$. In the same figure we also show the result for a Gaussian
potential. Clearly, the microscopic spectral density converges to the 
asymptotic result for $n\to \infty$. Both in  Fig. 3 and Fig. 4, 
the average spectral density $\rho(0)$ depends on $n$ because of the
normalization of our weight function.

\section{Conclusions}
\label{conclusions}

We have shown universality for the chiral Orthogonal Ensembles. Our proof is
based on a relation between the kernels for $\beta =1$ and $\beta =2$ and
the universality of the kernel for $\beta =2$. In this article we have
completed the proof outlined in \cite{Senerprl} by deriving
an asymptotic property of the skew-orthogonal polynomials which relates
an integral over the region near the largest zero to an integral in the
microscopic region. Universality now has been shown for all three chiral
ensembles. 

An alternative method for ensembles with $\beta = 1$ and $\beta = 4$
was recently proposed in \cite{Widom}. This method 
does not rely on the construction of 
the skew-orthogonal polynomials at all, but it is our 
point of view is that both methods are equivalent. For
example, the matrix elements of some of the operators in our construction
are also required in the method proposed by Widom. The advantage of our
method is best illustrated by the result of this article in which 
universality has been proved by means of  an asymptotic relation of 
skew-orthogonal polynomials. It would be interesting to identify this 
relation within Widom's  approach.

Finally, we wish to mention an alternative way of looking at universality.
For theories with broken chiral symmetry and  a mass gap we have two types
of modes, the soft modes and the hard modes. An effective partition function
is obtained by integrating out the hard modes.
If the Goldstone bosons corresponding to the spontaneous breaking
of chiral symmetry are the only soft modes,  
there is no need to do this calculation. The effective partition function 
can be written down solely on the basis of the
symmetries of the theory and is thus the same for all partition functions
with the same global symmetries. 
The equivalence
of the effective theory for the Goldstone modes and chiral Random Matrix Theory
has been demonstrated nonperturbatively for
the chiral Unitary Ensemble \cite{Vplb,Osbornprl,OTV,DOTV}. 
For the the other two values of the Dyson
index this has only been shown perturbatively \cite{Toub99}.

The mass of the  relevant 
Goldstone modes in the generating function of the spectrum is proportional
to the square root of the distance to $\lambda = 0$.  
Universal behavior is thus immanent in the microscopic limit. 
However, proving universality is 
equivalent to showing that the theory has a mass gap and that
the Goldstone modes are the only soft modes. In QCD this is equivalent
to proving confinement. 

\vskip 0.5 cm
{\bf Acknowledgements}
\vskip 0.5cm

This work was partially supported by the US DOE grant
DE-FG-88ER40388. Denis Dalmazi, Poul Damgaard, Melih \c Sener, 
Dominique Toublan  and Tilo Wettig are thanked for useful discussions.
J.J.M.V. is grateful to the Institute for Nuclear Theory at
the University of Washington for its hospitality and partial support
during the completion of this work.

\end{document}